\documentstyle[12pt]{article}

\begin{document}
\setcounter{page}{1} \pagestyle{plain} \vspace{1cm}
\begin{center}
\Large{\bf $\alpha$-Attractors and Reheating in a Non-Minimal Inflationary Model}\\
\small \vspace{1cm} {\bf R.
Shojaee$^{a,}$\footnote{shojaei.raheleh@gmail.com}},\quad {\bf K.
Nozari$^{b,}$\footnote{knozari@umz.ac.ir }}\quad and \quad
{\bf F. Darabi$^{a,}$\footnote{f.darabi@azaruniv.ac.ir}}\\
\vspace{0.5cm}
$^{a}$Department of Physics, Azarbaijan Shahid Madani University,\\
P. O. Box 53714-161, Tabriz, Iran\\
$^{b}$Department of Theoretical Physics, Faculty of Basic Sciences,\\
University of Mazandaran, P. O. Box 47416-95447, Babolsar, Iran\\
\end{center}
\vspace{1.5cm}

\begin{abstract}

We study $\alpha$-attractor models with both E-model and T-model
potential in an extended Non-Minimal Derivative (NMD) inflation
where a canonical scalar field and its derivatives are non-minimally
coupled to gravity. We calculate the evolution of perturbations
during this regime. Then by adopting inflation potentials of the
model we show that in the large $N$ and small $\alpha$ limit, value
of the scalar spectral index $n_{s}$ and tensor to scalar ratio $r$
are universal. Next, we study reheating after inflation in this
formalism. We obtain some constraints on the model's parameter space
by adopting the results with Planck 2018.

\end{abstract}

\vspace{1.5cm}
\newpage

\section{Introduction}

The big bang theory had a number of problems. Inflationary paradigm
has emerged as candidate for solve this problems that we can see
today [1-9]. Recently, the Planck measurements of temperature
fluctuation of cosmic microwave background (CMB) shows that spectrum
of primordial scalar perturbations has small deviation from scale
invariance. It is encoded in scalar spectral index in form
$n_{s}=0.9649\pm0.0042$. Furthermore, the Planck has placed a
constraint on the tensor to scalar ratio $r<0.056$. New inflationary
models provide compelling explanation of such perturbations. During
the last five years, a broad class of inflationary models have been
proposed. So that in the context of completely different theories,
all of which share the same predictions providing an excellent match
to the WMAP and Planck data [10-13]. These models are called
'Cosmological Attractors'. To see more details, one can refer to
Refs. [14-22].

There are several types of attractor models which among them we
study $\alpha$-attractors model [16-19]. $\alpha$-attractor scenario
has been proposed in the framework of supergravity, so that $\alpha$
describe moduli space curvature given by $R_{K}=\frac{2}{3\alpha}$.
This model provide a very good fit to the observational data. There
are two types of $\alpha$-attractor models called E-model and
T-model according to its potentials. In the large e-folds number and
small $\alpha$ limit, predicted for scalar spectral index and
tensor-to-scalar ratio, measured by CMB experiments, have universal
form as

\begin{equation}
n_{s}\approx1-\frac{2}{N}\,\,\,\,,\,\,\,\,r\approx\frac{12\alpha}{N^{2}}.
\end{equation}

For $\alpha=1$ is coincide with predictions of Higgs inflation and
Starobinsky model and for large $\alpha$ theory asymptotes to the
power-law inflation. (for more details on the issue of
$\alpha$-attractors refer to Refs. [23-25])

On the other hand, consideration of process by which the universe
reheats, provide additional constraints to inflationary models
[26-30]. Universe after inflation without reheating, would be cold
and empty. After inflation ends, inflaton begins to oscillate around
the minimum of its potential. When inflaton oscillates, energy
stored in inflaton field is transferred to the plasma of
relativistic particles.

Reheating era is an important part of inflationary models for
several reasons. Firstly, It can be explains how inflation is
connected to hot big bang phase. Also, micro physics of this era
depends on the interaction between inflaton and other fundamental
fields, then by constraining reheating, we can learn about coupling
this fields.

Recently, show that the Planck satellite measurements of CMB
anisotropies constrain kinematic properties of the reheating era. By
studying reheating process we are able to additional constrains on
inflationary models [31]. Reheating epoch characterized by three
major parameters: e-fold number $N_{re}$, reheating temperature
$T_{re}$ and effective equation of state parameter $\omega_{re}$.
Once inflaton field decays, radiation dominated epoch starts and the
universe obtains a temperature. Measuring this temperature is
important for understanding thermal history of our universe.
Furthermore, in standard reheating scenario effective equation of
state parameter is zero. But numerical experiments of thermalization
phase suggest a possibly wide range of values $0\leq\omega\leq0.25$
[32]. Though, $\omega_{re}>-\frac{1}{3}$ is is required at end of
inflation but $\omega_{re}<\frac{1}{3}$ is needed for beginning of
the radiation-dominated phase. This parameters also provides some
constraints to the models.

In this work we extend the nonminimal inflationary models to the
case that a canonical scalar field is coupled nonminimally to the
gravity and in the same time its derivatives are also coupled to the
Einstein's tensor. We study cosmological inflation to find possible
$\alpha$-attractor then we consider in detail reheating phase in
this nonminimal inflationary model. We compare results with latest
observational data to see the viability of this extended model. We
obtain some constraints on the model's parameter space.

\section{The Model}

In this letter, we study models with a canonical scalar field so
that gravity is coupled non-minimally to both the scalar field and
its derivatives [33-34]. The general action for such model described
by the following expression

\begin{equation}
S=\frac{1}{2}\int
d^{4}x\sqrt{-g}\Bigg[M_{p}^{2}f(\phi)R+\frac{1}{\widetilde{M}^{2}}G_{\mu\nu}\partial^{\mu}\phi\partial^{\nu}\phi-2V(\phi)\Bigg]\,,
\end{equation}

where $\phi$ is a scalar field, $f(\phi)$ is a general function of
$\phi$, $M_{p}$ is a reduced Planck mass, and $\widetilde{M}$ is a
mass parameter.

\subsection{Field Equations}

First let us assume that $f(\phi)=\frac{\phi^{2}}{2}$. In a
spatially flat FLRW geometry, the equations of motion following from
action (2) are [27]

\begin{equation}
H^{2}=\frac{1}{3M_{p}^{2}}\Bigg[-\frac{3}{2}M_{p}^{2}H\phi(2\dot{\phi}+H\phi)+\frac{9H^{2}}{2\widetilde{M}^{2}}\dot{\phi}^{2}+V(\phi)\Bigg]\,,
\end{equation}

\begin{eqnarray}\nonumber
\dot{H} &=&
-\frac{1}{2M_{p}^{2}}\Bigg[\dot{\phi}^{2}\bigg(\frac{3H^{2}}{\widetilde{M}^{2}}-\frac{\dot{H}}{\widetilde{M}^{2}}\bigg)
-\frac{2H}{\widetilde{M}^{2}}\dot{\phi}\ddot{\phi}-\frac{3}{2}M_{p}^{2}H\phi(2\dot{\phi}+H\phi)
\\
&&
+\frac{1}{2}M_{p}^{2}\bigg((2\dot{H}+3H^{2})\phi^{2}+4H\phi\dot{\phi}+2\phi\ddot{\phi}+2\dot{\phi}^{2}\bigg)\Bigg]\,,
\end{eqnarray}

\begin{equation}
-3M_{p}^{2}(2H^{2}+\dot{H})\phi+\frac{3H^{2}}{\widetilde{M}^{2}}\ddot{\phi}+3H\bigg(\frac{3H^{2}}{\widetilde{M}^{2}}
+\frac{2\dot{H}}{\widetilde{M}^{2}}\bigg)\dot{\phi}+V'(\phi)=0\,,
\end{equation}

where a dot denotes derivative with respect to the cosmic time and
prime refers  to  derivative with respect to the scalar field. The
slow-roll parameters are defined as

\begin{equation}
\epsilon\equiv-\frac{\dot{H}}{H^{2}}\,\,\,\,,\,\,\,\,\eta\equiv-\frac{1}{H}\frac{\ddot{H}}{\dot{H}}\,.
\end{equation}

In this regard, we find these parameters as follows

\begin{equation}
\epsilon=\bigg[1+\frac{\phi^{2}}{2}-\frac{\dot{\phi^{2}}}{2\widetilde{M}^{2}M_{p}^{2}}\bigg]^{-1}
\bigg[\frac{3\dot{\phi}^{2}}{2\widetilde{M}^{2}M_{p}^{2}}+\frac{\phi\dot{\phi}}{2H}
+\frac{\ddot{\phi}}{H\dot{\phi}}\bigg(\frac{\phi\dot{\phi}}{2H}
-\frac{\dot{\phi^{2}}}{\widetilde{M}^{2}M_{p}^{2}}\bigg)\bigg]\,,
\end{equation}

\begin{equation}
\eta=-2\epsilon-\frac{\dot{\epsilon}}{H\epsilon}\,.
\end{equation}

To have inflation phase, should satisfy the slow-roll conditions
($\epsilon\ll1$ , $\eta\ll1$). In our setup within the slow-roll
approximation, the equations of motion (3-5) is obtained as follows
respectively

\begin{equation}
H^{2}\simeq\frac{1}{3M_{p}^{2}}\Bigg[-\frac{3}{2}M_{p}^{2}H^{2}\phi^{2}+V(\phi)\Bigg]\,,
\end{equation}

\begin{equation}
\dot{H}\simeq-\frac{1}{2M_{p}^{2}}\Bigg[\frac{3H^{2}\dot{\phi}^{2}}{\widetilde{M}^{2}}-M_{p}^{2}H\phi\dot{\phi}+M_{p}^{2}\dot{H}\phi^{2}\Bigg]\,,
\end{equation}

\begin{equation}
-6M_{p}^{2}H^{2}\phi+\frac{9H^{3}\dot{\phi}}{\widetilde{M}^{2}}+V'(\phi)\simeq0\,.
\end{equation}

In this respect, within the slow-roll conditions, (7) become

\begin{equation}
\epsilon\simeq\frac{\widetilde{M}^{2}M_{p}^{4}}{2V}\Bigg[[(1+\frac{\phi^{2}}{2})(\frac{V'}{V})]^{2}
-\frac{3\phi}{M_{p}^{2}}(1+\frac{\phi^{2}}{2})(\frac{V'}{V})+2\phi^{2}\Bigg]\,.
\end{equation}

The e-folds number during inflation is defined as

\begin{equation}
{\cal N}=\int_{t_{hc}}^{t_{e}}H\,dt\,,
\end{equation}

where $t_{hc}$ and $t_{e}$ are defined as the time of horizon
crossing and end of inflation, respectively
($\epsilon(t_{hc})=\epsilon(t_{e})=1$). In our setup, the e-fold
number in the slow-roll approximation can be written as follows

\begin{equation}
{\cal
N}\simeq\int_{\phi_{hc}}^{\phi_{e}}\frac{V(\phi)d\phi}{M_{p}^{2}\bigg(1
+\frac{1}{2}\phi^{2}\bigg)\Bigg[2M_{p}^{2}\widetilde{M}^{2}\phi
-M_{p}^{2}\widetilde{M}^{2}\frac{V'(\phi)}{V(\phi)}\bigg(1+\frac{1}{2}\phi^{2}\bigg)\Bigg]}\,.
\end{equation}

\subsection{Linear Perturbations}

Now, we study linear perturbations of the model around the
homogeneous background solution. In the first step, we should expand
the action (2) up to the second order in small fluctuations of the
space-time background metric. It is convenient to work in
Arnowitt-Deser-Misner (ADM) formalism in which one can eliminate one
extra degree of freedom of the perturbations from the beginning. In
this formalism the space-time metric is given by [35]

\begin{equation}
ds^{2}=-N^{2}dt^{2}+h_{ij}(N^{i}dt+dx^{i})(N^{j}dt+dx^{j})\,,
\end{equation}

where $N^{i}$ and $N$ are the shift vector and lapse function,
respectively. We can obtain general perturbed form of the ADM metric
by expanding $N^{i}$ as $N^{i}=\delta^{ij}\partial_{j}\Upsilon$ and
$N$ as $N=1+2\Phi$ in which $\Phi$ and $\Upsilon$ are three-scalars.
Also, the coefficient $h_{ij}$ in the second term of the above
metric can be written as
$h_{ij}=a^{2}(t)[(1+2\Psi)\delta_{ij}+\gamma_{ij}]$, where $\Psi$ is
spatial curvature perturbation and $\gamma_{ij}$ is symmetric and
traceless shear three-tensors. In the rest, we choose
$\gamma_{ij}=0$ and $\delta\Phi=0$. Finally, we can write the
perturbed metric up to the linear level as [36]

\begin{equation}
ds^{2}=-(1+2\Phi)dt^{2}+2\partial_{i}\Upsilon
dtdx^{i}+a^{2}(t)(1+2\Psi)\delta_{ij}dx^{i}dx^{j}\,.
\end{equation}

By replacing the perturbed metric in action (2) and expanding it up
to the second order in small perturbations, the second-order action
can be rewritten as follows [37]

$$S^{(2)}=\int dt dx^{3}a^{3}\Bigg[-\frac{3}{2}(M_{p}^{2}\phi^{2}-\frac{\dot{\phi}^{2}}{\widetilde{M}^{2}})\dot{\Psi}^{2}
+\frac{1}{a^{2}}((M_{p}^{2}\phi^{2}-\frac{\dot{\phi}^{2}}{\widetilde{M}^{2}})\dot{\Psi}$$
$$-(M_{p}^{2}H\phi^{2}+M_{p}^{2}\phi\dot{\phi}-\frac{3H\dot{\phi}^{2}}{\widetilde{M}^{2}})\Phi)\partial^{2}\Upsilon
-\frac{1}{a^{2}}(M_{p}^{2}\phi^{2}-\frac{\dot{\phi}^{2}}{\widetilde{M}^{2}})\Phi\partial^{2}\Psi$$
$$+3(M_{p}^{2}H\phi^{2}+M_{p}^{2}\phi\dot{\phi}-\frac{3H\dot{\phi}^{2}}{\widetilde{M}^{2}})\Phi\dot{\Psi}
+3H(-\frac{1}{2}M_{p}^{2}H\phi^{2}-M_{p}^{2}\phi\dot{\phi}$$
\begin{equation}
+\frac{3H\dot{\phi}^{2}}{\widetilde{M}^{2}})\Phi^{2}
+\frac{1}{2a^{2}}(M_{p}^{2}\phi^{2}+\frac{\dot{\phi}^{2}}{\widetilde{M}^{2}})(\partial\Psi)^{2}\Bigg]\,.
\end{equation}

By variation of the above action with respect to $N^{i}$ and $N$ we
obtain

\begin{equation}
\Phi=\frac{M_{p}^{2}\phi^{2}-\frac{\dot{\phi}^{2}}{\widetilde{M}^{2}}}{M_{p}^{2}H\phi^{2}+M_{p}^{2}\phi\dot{\phi}-
\frac{3H\dot{\phi}^{2}}{\widetilde{M}^{2}}}\dot{\Psi}\,,
\end{equation}

$$\partial^{2}\Upsilon=\frac{2a^{2}}{3}\frac{(-\frac{9}{2}M_{p}^{2}H^{2}\phi^{2}-9M_{p}^{2}H\phi\dot{\phi}
+\frac{27H^{2}\dot{\phi}^{2}}{\widetilde{M}^{2}})}{(M_{p}^{2}H\phi^{2}+M_{p}^{2}\phi\dot{\phi}-\frac{3H\dot{\phi}^{2}}{\widetilde{M}^{2}})}$$
\begin{equation}
+3\dot{\Psi}a^{2}-\frac{M_{p}^{2}\phi^{2}-\frac{\dot{\phi}^{2}}{\widetilde{M}^{2}}}{M_{p}^{2}H\phi^{2}+M_{p}^{2}\phi\dot{\phi}-
\frac{3H\dot{\phi}^{2}}{\widetilde{M}^{2}}}\dot{\Psi}\,.
\end{equation}

By substituting the equation (18) in equation (17) and taking some
integration by part, the quadratic action can be rewritten as
following expression

\begin{equation}
S^{(2)}=\int dt
dx^{3}a^{3}\vartheta_{s}\bigg[\dot{\Psi}^{2}-\frac{c_{s}^{2}}{a^{2}}(\partial\Psi)^{2}\bigg]
\end{equation}

where the parameters $\vartheta_{s}$ and $c_{s}^{2}$ are expressed
as

\begin{equation}
\vartheta_{s}\equiv6\frac{(M_{p}^{2}\phi^{2}-\frac{\dot{\phi}^{2}}{\widetilde{M}^{2}})^{2}(-\frac{1}{2}M_{p}^{2}H^{2}\phi^{2}-M_{p}^{2}H\phi\dot{\phi}+\frac{3}{\widetilde{M}^{2}}
H^{2}\dot{\phi}^{2})}{(M_{p}^{2}H\phi^{2}+M_{p}^{2}\phi\dot{\phi}-\frac{3}{\widetilde{M}^{2}}H\dot{\phi}^{2})^{2}}+
3\bigg(\frac{1}{2}M_{p}^{2}\phi^{2}-\frac{1}{2\widetilde{M}}\dot{\phi}^{2}\bigg)\,,
\end{equation}

and

\begin{eqnarray}\nonumber
c_{s}^{2} &\equiv&
\frac{3}{2}\bigg\{\bigg(M_{p}^{2}\phi^{2}-\frac{\dot{\phi}^{2}}{\widetilde{M}^{2}}\bigg)^{2}
\bigg(M_{p}^{2}H\phi^{2}+M_{p}^{2}\phi\dot{\phi}-\frac{3H\dot{\phi}^{2}}{\widetilde{M}^{2}}\bigg)H
-4\bigg(M_{p}^{2}H\phi^{2}+M_{p}^{2}\phi\dot{\phi}-\frac{3H\dot{\phi}^{2}}{\widetilde{M}^{2}}\bigg)^{2}
\\
 \nonumber && \bigg(M_{p}^{2}\phi^{2}-\frac{\dot{\phi}^{2}}{\widetilde{M}^{2}}\bigg)\bigg(M_{p}^{2}\phi^{2}-\frac{\dot{\phi}^{2}}{\widetilde{M}^{2}}\bigg)
 \bigg(M_{p}^{2}\phi\dot{\phi}-\frac{\dot{\phi}\ddot{\phi}}{\widetilde{M}^{2}}\bigg)
\bigg(M_{p}^{2}H\phi^{2}+M_{p}^{2}\phi\dot{\phi}-\frac{3H\dot{\phi}^{2}}{\widetilde{M}^{2}}\bigg)
\\
 \nonumber && -\bigg(M_{p}^{2}-\frac{\dot{\phi}^{2}}{\widetilde{M}^{2}}\bigg)^{2}\bigg(M_{p}^{2}\dot{H}\phi^{2}+
 2M_{p}^{2}H\phi\dot{\phi}M_{p}^{2}\dot{\phi}^{2}+M_{p}^{2}\phi\ddot{\phi}
-\frac{3\dot{H}\dot{\phi}^{2}}{\widetilde{M}^{2}}-\frac{6}{\widetilde{M}^{2}}H\dot{\phi}\ddot{\phi}\bigg)\bigg\}
\\
 \nonumber && \bigg\{9\bigg[\frac{1}{2}M_{p}^{2}\phi^{2}-\frac{\dot{\phi}^{2}}{2\widetilde{M}^{2}}\bigg]\bigg[4\bigg(\frac{1}{2}M_{p}^{2}\phi^{2}
 -\frac{\dot{\phi}^{2}}{2\widetilde{M}^{2}}\bigg)
\bigg(-\frac{1}{2}M_{p}^{2}H^{2}\phi^{2}-M_{p}^{2}H\phi\dot{\phi}+\frac{3}{\widetilde{M}^{2}H^{2}\dot{\phi}^{2}}\bigg)
\\
 && +\bigg(M_{p}^{2}H\phi^{2}+M_{p}^{2}\phi\dot{\phi}-\frac{3H\dot{\phi}^{2}}{\widetilde{M}^{2}}\bigg)^{2}\bigg]\bigg\}^{-1}\,.
\end{eqnarray}

Let us now calculate quantum perturbations of $\Psi$. To this end,
we can obtain equation of motion of the curvature perturbation by
variation of action (20) which follows

\begin{equation}
\ddot{\Psi}+\bigg(3H+\frac{\dot{\vartheta_{s}}}{\vartheta_{s}}\bigg)\dot{\Psi}+\frac{c_{s}^{2}k^{2}}{a^{2}}\Psi=0\,.
\end{equation}

By solving the equation of motion up to the lowest order of the
slow-roll approximation, we obtain

\begin{equation}
\Psi=\frac{iH\exp(-ic_{s}k\tau)}{2c_{s}^{\frac{3}{2}}\sqrt{k^{3}}\vartheta_{s}}(1+ic_{s}k\tau)\,.
\end{equation}

Two-point correlation function of curvature perturbations can be
found by obtaining vacuum expectation value at the end of inflation.
By using this function, we can study power spectrum of the curvature
perturbation in our model. We describe the power spectrum $P_{s}$ as
following expression

\begin{equation}
\langle0|\Psi(0,\textbf{k}_{1})\Psi(0,\textbf{k}_{2})|0\rangle=\frac{2\pi^{2}}{k^{3}}P_{s}(2\pi)^{3}\delta^{3}(\textbf{k}_{1}+\textbf{k}_{2})\,,
\end{equation}

where by definition

\begin{equation}
P_{s}=\frac{H^{2}}{8\pi^{2}\vartheta_{s} c_{s}^{3}}\,.
\end{equation}

The spectral index is given by [38-42]

\begin{equation}
n_{s}-1=\frac{d\ln P_{s}}{d\ln
k}|_{c_{s}k=aH}=-2\epsilon-\delta_{F}-\eta_{s}-S,
\end{equation}

by definition

\begin{equation}
\delta_{F}=\frac{\dot{f}}{H(1+f)}\,\,\,\,,\,\,\,\,\eta_{s}=\frac{\dot{\epsilon_{s}}}{H\epsilon_{s}}\,\,\,\,,\,\,\,\,S=\frac{\dot{c_{s}}}{Hc_{s}},
\end{equation}

where

\begin{equation}
\epsilon_{s}=\frac{\vartheta_{s}c_{s}^{2}}{M_{p}^{2}(1+f)},
\end{equation}

which gives finally

\begin{equation}
n_{s}-1=-2\epsilon-\frac{1}{H}\frac{d\ln c_{s}}{dt}
-\frac{1}{H}\frac{d\ln[2H(1+\frac{\phi^{2}}{2})\epsilon+\phi\dot{\phi}]}{dt}\,,
\end{equation}
which shows the scale dependence of perturbations due to deviation
of $n_{s}$ from $1$.

Let us now study the amplitude of tensor perturbation and its
spectral index. In this regard, we consider the metric as follows

\begin{equation}
ds^{2}=-dt^{2}+a(t)^{2}(\delta_{ij}+T_{ij})dx^{i}dx^{j}\,,
\end{equation}

where $T_{ij}$ is transverse and traceless shear three-tensor. We
set $T_{ij}$ in terms of the two polarization tensors, as follows

\begin{equation}
T_{ij}=T_{+}e^{+}_{ij}+T^{\times}e^{\times}_{ij}\,,
\end{equation}

where $e^{+}_{ij}$ and $e^{\times}_{ij}$ are symmetric, transverse
and traceless. In this case, the second-order action for the tensor
mode of the perturbations (known as gravitational waves) can be
written as

\begin{equation}
S_{T}=\int dt dx^{3}
a^{3}\vartheta_{T}\bigg[\dot{T}_{(+,\times)}^{2}-\frac{c_{T}^{2}}{a^{2}}(\partial
T_{(+,\times)})^{2}\bigg]\,,
\end{equation}

by definition

\begin{equation}
\vartheta_{T}\equiv\frac{1}{8}(M_{p}^{2}\phi^{2}-\frac{\dot{\phi}^{2}}{\widetilde{M}^{2}})
\end{equation}
and
\begin{equation}
c_{T}^{2}\equiv\frac{\widetilde{M}^{2}M_{p}^{2}\phi^{2}+\dot{\phi}^{2}}{\widetilde{M}^{2}M_{p}^{2}\phi^{2}-\dot{\phi}^{2}}\,.
\end{equation}

In our setup, the amplitude of the tensor perturbations is given by

\begin{equation}
P_{T}=\frac{H^{2}}{2\pi^{2}\vartheta_{T}c_{T}^{3}}\,,
\end{equation}

we have defined the tensor spectral index of the gravitational waves
as

\begin{equation}
n_{T}\equiv\frac{d\ln P_{T}}{d\ln
k}|_{c_{T}k=aH}\,=-2\epsilon-\delta_{F}.
\end{equation}

leads to the following expression

\begin{equation}
n_{T}=-2\epsilon-\frac{\phi\dot{\phi}}{H(1+\frac{\phi^{2}}{2})}\,.
\end{equation}

Finally, the ratio between the amplitudes of the tensor and scalar
perturbations (tensor-to-scalar ratio) in our setup is given by

\begin{equation}
r=\frac{P_{T}}{P_{s}}=16
c_{s}\bigg(\epsilon+\frac{\phi\dot{\phi}}{2H(1+\phi^{2}/2)}+o(\epsilon^{2})\bigg)\simeq-8c_{s}n_{T}.
\end{equation}

As we can see, NMD inflation yields the standard consistency
relation.

\section{$\alpha$-Attractors}

We consider a class of NMD inflationary models of
$\alpha$-attractors in a unified manner. As shown in Ref. [12]
$\alpha$-attractors are consistent with observations of the CMB
anisotropies. For consistency with observed $r$, the parameter
$\alpha$ should be less than $O(100)$. In this section, we focus on
the E-model and T-model as generalised models of $\alpha$-attractors
so that specified by the potentials as (for more details refer to
Refs.[14,17])

\begin{equation}
V=V_{0}\bigg[1-exp(-\sqrt{\frac{2}{3M_{pl}^{2}\alpha}}\phi)\bigg]^{2n},
\end{equation}

\begin{equation}
V=V_{0}\tanh^{2n}(\frac{\phi}{\sqrt{6M_{pl}^{2}\alpha}}),
\end{equation}

respectively. $V_{0}$ and $n$ are the parameters.

Now, by using the potentials defined in Eqs. (40) and (41) and
solving integral (14), we find the value of the field at the horizon
crossing. We can therefore substitute this value in Eqs. (12), (30)
and (39). Finally, when $n=1$ and $\alpha\ll1$, also using the
slow-roll approximation, we can find the values of the scalar
spectral index and the tensor-to-scalar ratio for each model. Under
these conditions, the values of $n_{s}$ and $r$ for E-model
potential are given by

\begin{equation}
n_{s}=1-\frac{2}{1+N+3\alpha\ln(1+N)/4},
\end{equation}

\begin{equation}
r=\frac{12\alpha}{(1+N)^{2}+3\alpha(1+N)\ln(1+N)/2},
\end{equation}

where in the large e-folds number and small $\alpha$ limit become

\begin{equation}
n_{s}\approx1-\frac{2}{N},
\end{equation}

\begin{equation}
r\approx\frac{12\alpha}{N^{2}}.
\end{equation}

As we can see, predicted form of the scalar spectral index and
tensor-to-scalar ratio have universal form.

Also, Under above conditions, the values of $n_{s}$ and $r$ for
T-model potential can be written as follows

\begin{equation}
n_{s}=1-\frac{2}{N}+\frac{2}{N^{2}+N},
\end{equation}

\begin{equation}
r=\frac{12\alpha}{(1+N)^{2}}.
\end{equation}

Now, we study the perturbation parameters numerically. To this end,
we plot tensor to scalar ratio versus the scalar spectral index and
compare them with the observations. The results are shown in Figure
(1). As we shown, the models for $N=50, 60$ and $70$ are consistent
with the Planck measurements.

\begin{figure}
\begin{center}\includegraphics{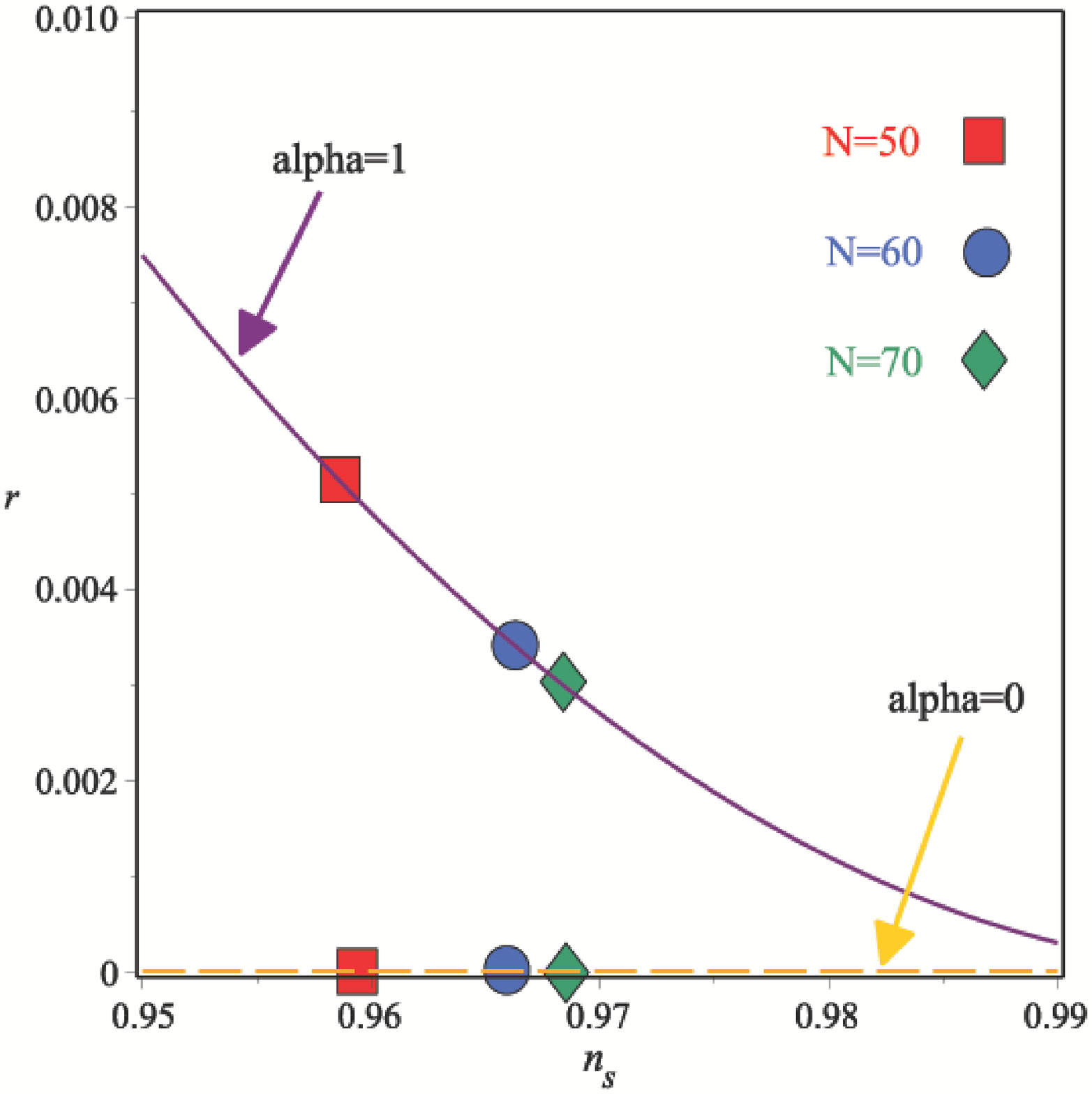} \vspace{5.5cm}\includegraphics{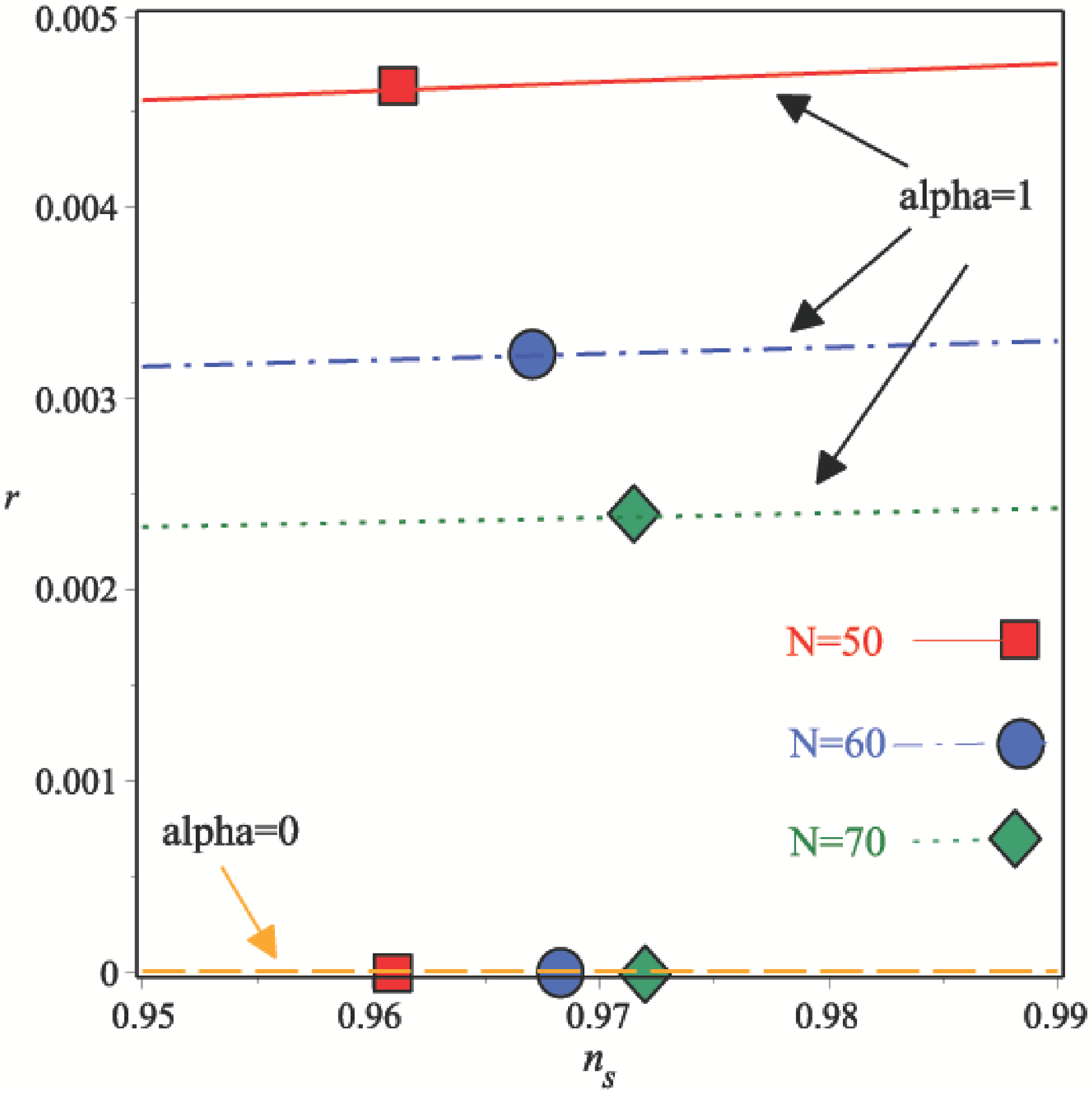}
\end{center}
\caption{\small {Tensor to scalar ratio versus the scalar spectral
index for NMD model with the E-model (left panel) and T-model (right
panel) potential.}}
\end{figure}

\section{Reheating}

After inflation ends, the process of reheating take places. By
consider this process we can investigate some constraints on the
model's parameter space. In this section, we obtain expressions for
e-fold number and temperature of reheating era.

The e-fold number between the time of horizon crossing and end of
inflation is written as

\begin{equation}
N_{hc}=\ln\bigg(\frac{a_{e}}{a_{hc}}\bigg),
\end{equation}

where $a_{hc}$ and $a_{e}$ are scale factor at the horizon crossing
and end of inflation, respectively.

We assume that at the end of reheating epoch, energy density of the
universe is given by

\begin{equation}
\rho_{re}=\frac{\pi^{2}g_{re}}{30}T_{re}^{4},
\end{equation}

where $g_{re}$ is effective number of relativistic particles at the
end of reheating.

In this respect, the length of reheating is written as

\begin{equation}
N_{re}=\ln\bigg(\frac{a_{re}}{a_{e}}\bigg)=-\frac{1}{3(1+\omega_{eff})}\ln
\bigg(\frac{\rho_{re}}{\rho_{e}}\bigg),
\end{equation}

so that we substituted the relation $\rho\sim
a^{-3(1+\omega_{eff})}$ at above equation. Where $\rho$ and
$\omega_{eff}$ are energy density and effective equation of state,
respectively.

Horizon crossing occurs at $a_{hc}H_{hc}=k$, where $k$ denotes wave
number and $H_{hc}$ is Hubble parameter at horizon crossing during
inflation. Then we can write

\begin{equation}
0=\ln\bigg(\frac{k}{a_{hc}H_{hc}}\bigg)=\ln\bigg(\frac{a_{e}}{a_{hc}}\frac{a_{re}}{a_{e}}\frac{a_{0}}{a_{re}}\frac{k}{a_{0}H_{hc}}\bigg),
\end{equation}

where $a_{0}$ is scale factor of present epoch. Now, from Eqs. (48)
and (50) we obtain

\begin{equation}
N_{hc}+N_{re}+\ln\bigg(\frac{a_{0}}{a_{re}}\bigg)
+\ln\bigg(\frac{k}{a_{0}H_{hc}}\bigg)=0.
\end{equation}

From conservation of entropy we can write

\begin{equation}
\frac{a_{0}}{a_{re}}=\bigg(\frac{43}{11g_{re}}\bigg)^{-\frac{1}{3}}\frac{T_{re}}{T_{0}},
\end{equation}

where $T_{0}$ is the current temperature. From Eqs. (49) and (53) we
obtain

\begin{equation}
\frac{a_{0}}{a_{re}}=\bigg(\frac{43}{11g_{re}}\bigg)^{-\frac{1}{3}}\bigg(\frac{\pi^{2}g_{re}}{30\rho_{re}}\bigg)^{-\frac{1}{4}}\frac{1}{T_{0}}.
\end{equation}

In our framework, the energy density during the inflation epoch can
be rewritten as follows

\begin{equation}
\rho=\frac{9H^{2}}{2\widetilde{M}^{2}}\dot{\phi}^{2}-\frac{3}{2}M_{p}^{2}H\phi(2\dot{\phi}+H\phi)+V(\phi).
\end{equation}

Now by setting $\epsilon=1$, the energy density at the end of
inflation phase is the following form

\begin{equation}
\rho_{e}=\bigg(\frac{2}{2+\phi_{e}^{2}}\bigg)V_{e},
\end{equation}

where $\phi_{e}$ and $V_{e}$ are scalar field and potential at the
end of inflation, respectively. Then from Eqs. (50) and (56) we have

\begin{equation}
\rho_{re}=\bigg(\frac{2}{2+\phi_{e}^{2}}\bigg)V_{e}\exp[-3(1+\omega_{eff})N_{re}].
\end{equation}

$H_{hc}$ can be find from Eq. (26). Now by substituting $H_{hc}$ and
Eqs. (54) and (57) in the Eq. (52) we can write the following
expression for e-fold number of reheating era

\begin{eqnarray}\nonumber
N_{re} &=&
\frac{4}{(1-3\omega_{eff})}\bigg[-N_{hc}-\ln\bigg(\frac{k_{hc}}{a_{0}}\bigg)+\frac{1}{2}\ln(8\pi^{2}P_{s}\vartheta_{s}c_{s}^{3})+
\frac{1}{3}\ln\bigg(\frac{43}{11g_{re}}\bigg) \\
 && +\frac{1}{4}\ln\bigg(\frac{\pi^{2}g_{re}}{30}\bigg)+\ln T_{0}-
\frac{1}{4}\ln\bigg(\frac{2V_{e}}{2+\phi_{e}^{2}}\bigg)\bigg].
\end{eqnarray}

and from equations (49) and (57) we obtain the temperature during
reheating

\begin{equation}
T_{re}=\bigg(\frac{60V_{e}}{\pi^{2}g_{re}(2+\phi_{e}^{2})}\bigg)^{\frac{1}{4}}\exp\bigg[-\frac{3}{4}(1+\omega_{eff})N_{re}\bigg].
\end{equation}

The scalar spectral index and wave number are related through $\phi$
with Eqs. (30), (48) and $a_{hc}H_{hc}=k$, therefore one can
investigate $N_{re}$ and $T_{re}$ in terms of the scalar spectral
index for each model.

\subsection{E-model}

First, we consider E-model potential defined in Eq. (40). As we
know, in the small $\alpha$ and large $N$ limit, one can rewrite the
potential as

\begin{equation}
V=V_{0}\bigg[1-2n\exp(-\sqrt{\frac{2}{3M_{pl}^{2}\alpha}}\phi)\bigg],
\end{equation}

where we assume that $n=1$. We also need to the e-fold number that
happen after a mode with wave number $k$. By substituting the above
equation in the Eq. (14) we can obtain e-folding as function of
$\phi$ which defined as following expression

\begin{eqnarray}\nonumber
N_{hc}&\approx&
\frac{V_{0}}{M_{pl}^{4}\widetilde{M}^{2}}\bigg[-\frac{1}{4}\phi^{2}(\phi^{2}-4)+\frac{1}{2}
(\phi^{4}+4\sqrt{6M_{pl}^{2}\alpha}\phi\\
&&
+12M_{pl}^{2}\alpha-4)\exp\bigg(-\sqrt{\frac{2}{3M_{pl}^{2}\alpha}}\phi\bigg)\bigg]_{\phi_{hc}}^{\phi_{e}}.
\end{eqnarray}

Now, by plotting the e-fold number and temperature of reheating era
versus the scalar spectral index we can study the parameters
numerically. We consider the case with $\alpha=1$ for some values of
effective equation of state. The results are shown in Figure (2).

\begin{figure}
\begin{center}\includegraphics{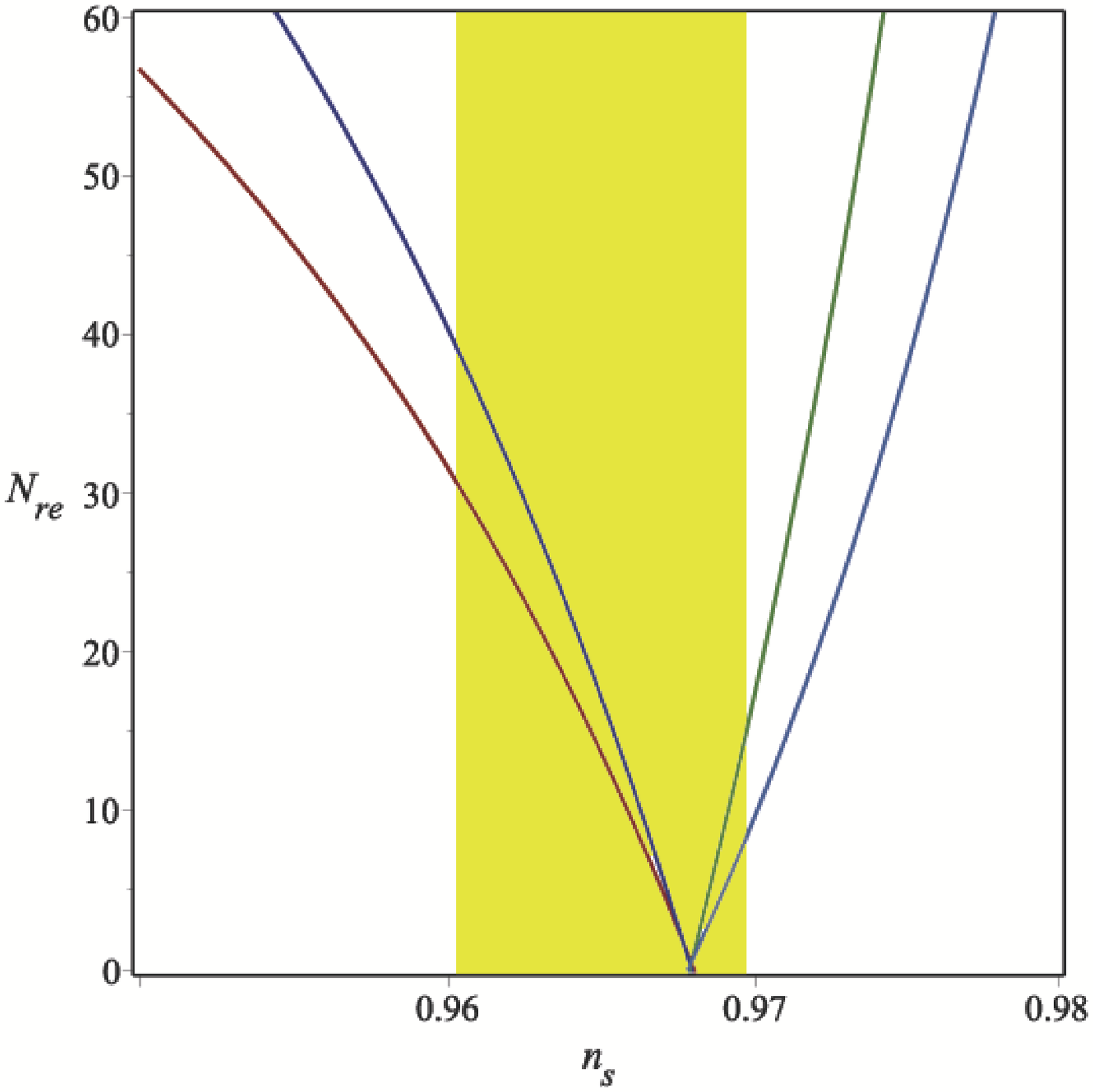} \vspace{5.5cm}\includegraphics{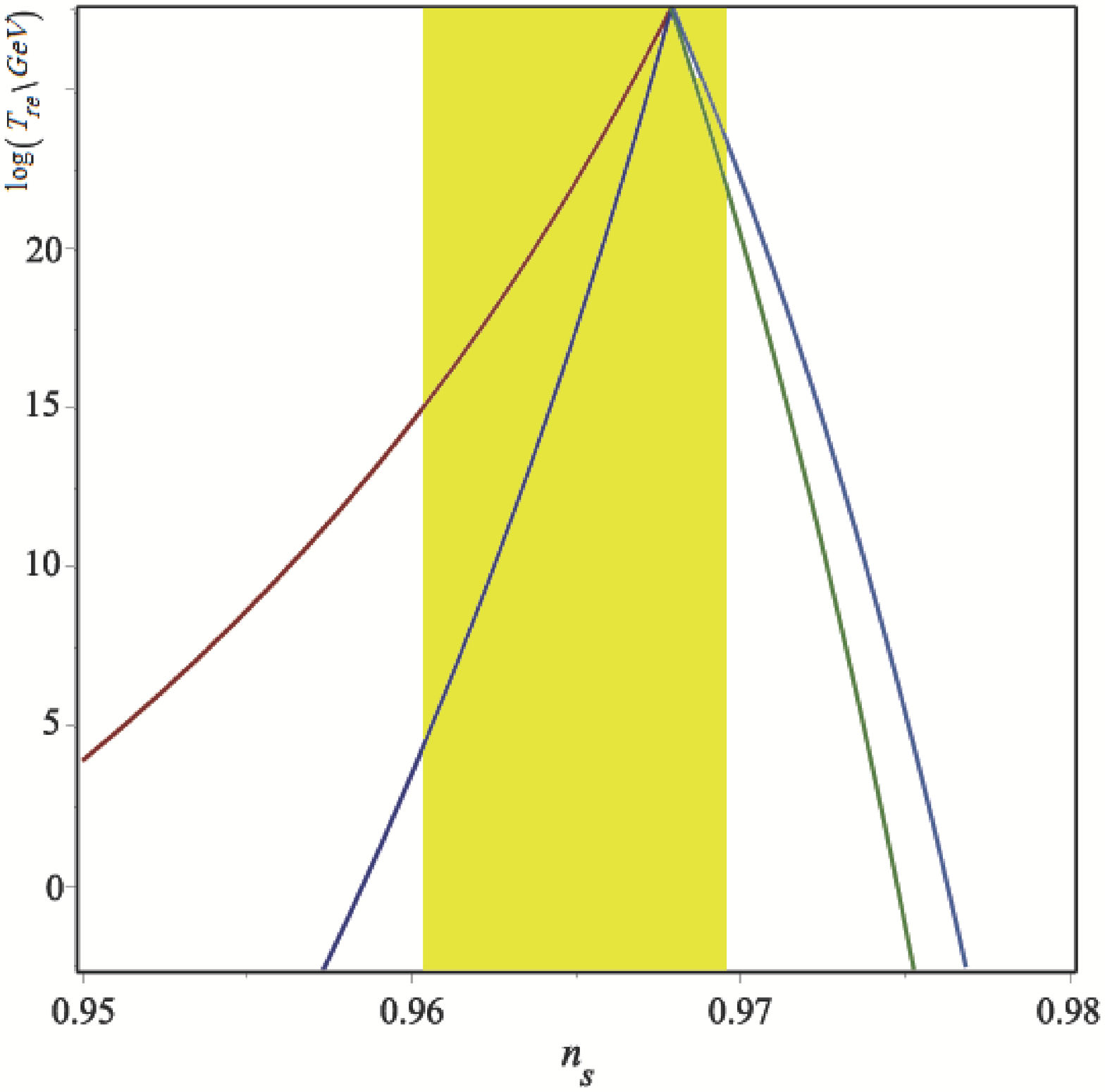}\vspace{1cm}
\end{center}
\caption{\small {$N_{re}$ (left panel) and $T_{re}$ (right panel) as
function of the scalar spectral index for E-model when $n=1$ with
$\alpha=1$. In each panel, the red line corresponds to
$\omega_{re}=-\frac{1}{3}$, the purple line to $\omega_{re}=0$, the
green line to $\omega_{re}=\frac{2}{3}$ and the blue line to
$\omega_{re}=1$. The yellow shaded region indicate the Planck 2018
constraint.}}
\end{figure}

According to the picture, we can obtain constraints on the model's
parameter space by adopting our results with the Planck 2018. The
constrains on the E-model parameters are shown in Table 1.

\begin{table}[ht]
\caption{The ranges of e-fold number and temperature during
reheating epoch for E-model potential which are consistent with the Planck 2018.}\vspace{0.5cm} % title of Table
\centering % used for centering table
\begin{tabular}{c c c c c} % centered columns (5 columns)
\hline\hline %inserts double horizontal lines
$n=1,\alpha=1$ & $\omega_{eff}=-\frac{1}{3}$ & $\omega_{eff}=0$ & $\omega_{eff}=\frac{2}{3}$ & $\omega_{eff}=1$ \\ [0.5ex] % inserts table
%heading
\hline % inserts single horizontal line
$N_{re}$ & $<31$ & $<39$ & $<15$ & $<8$ \\ % inserting body of the table
$\log(T_{re})$ & $>15$ & $>4$ & $>22$ & $>23$ \\ [1ex] % [1ex] adds vertical space
\hline %inserts single line
\end{tabular}
\label{table:nonlin} % is used to refer this table in the text
\end{table}

\subsection{T-model}

Now, we consider T-model potential defined in Eq. (41). By
substituting the potential in the Eq. (14) we can obtain e-folding
as function of $\phi$ which defined as following expression

\begin{eqnarray}\nonumber
N_{hc}&\approx&
V_{0}\bigg[12M_{pl}^{2}\alpha\log(\cosh(\frac{\phi}{\sqrt{6M_{pl}^{2}\alpha}}))+\frac{1}{4}(\phi^{4}-4)\cosh(\frac{\phi}{\sqrt{6M_{pl}^{2}\alpha}})^{-2}
\\
&&
-8\sqrt{6M_{pl}^{2}\alpha}\phi\tanh(\frac{\phi}{\sqrt{6M_{pl}^{2}\alpha}})-\phi^{4}+4\phi^{2}\bigg]_{\phi_{hc}}^{\phi_{e}}.
\end{eqnarray}

We consider $\alpha=1$ and plot the e-fold number and temperature of
reheating era versus the scalar spectral index. The results are
shown in Figure (3). Therefore, we obtain some constraints on the
model's parameter space. The constrains on the T-model parameters
are shown in Table 2.

\begin{figure}
\begin{center}\includegraphics{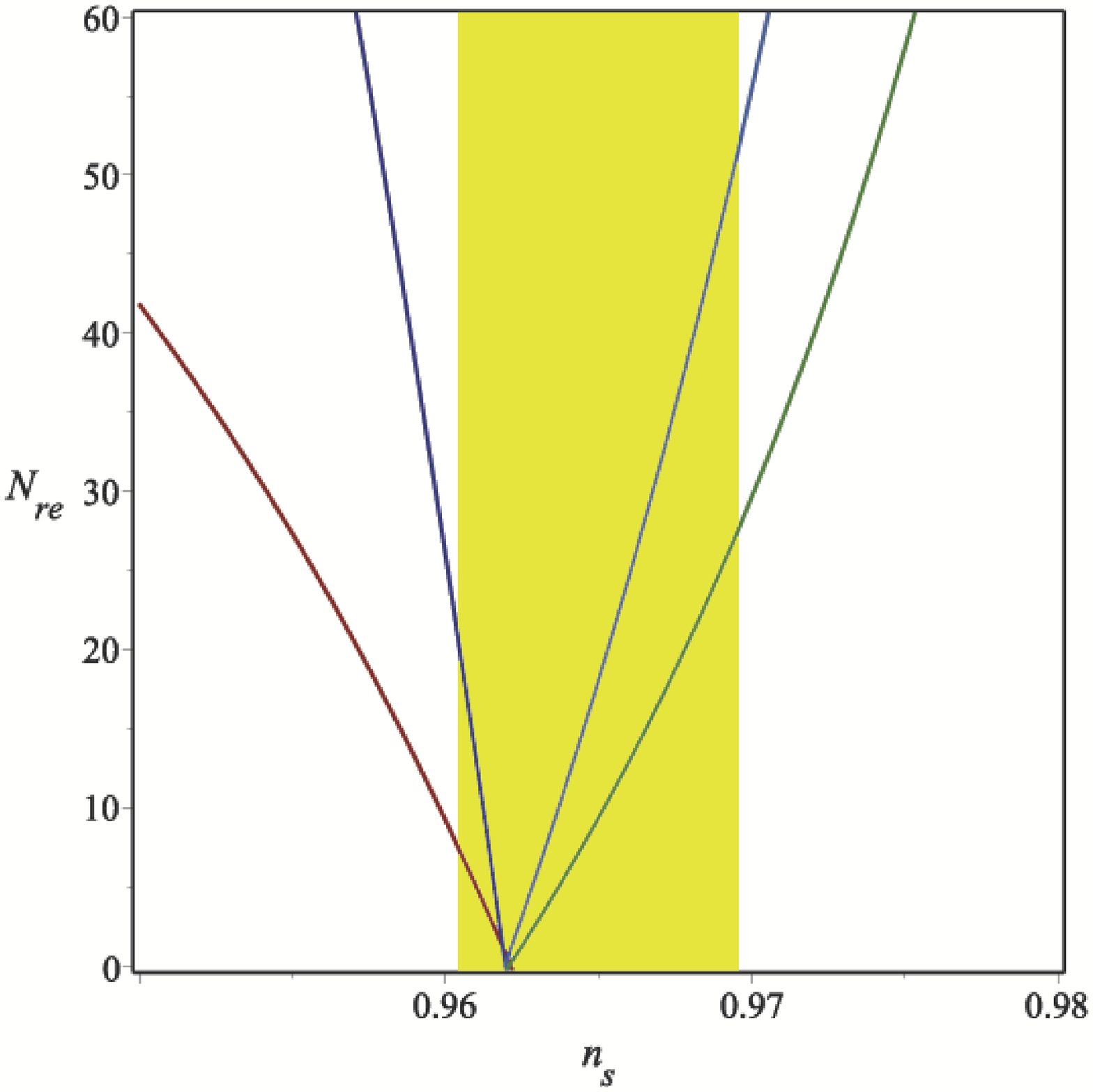} \vspace{5.5cm}\includegraphics{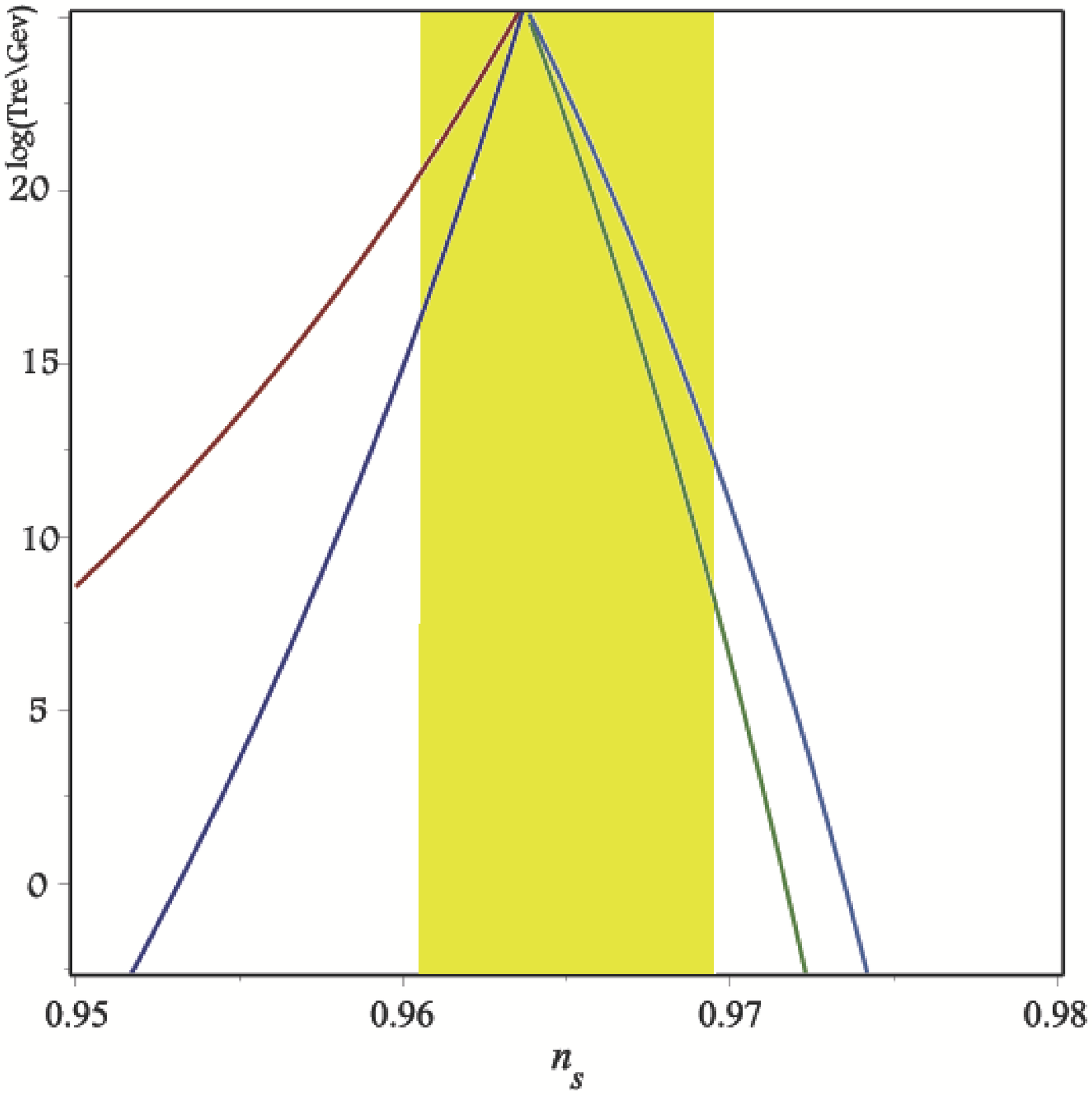}\vspace{1cm}
\end{center}
\caption{\small {$N_{re}$ (left panel) and $T_{re}$ (right panel) as
function of scalar spectral index for T-model when $n=1$
with$\alpha=1$. In each panel, the red line corresponds to
$\omega_{re}=-\frac{1}{3}$, the purple line to $\omega_{re}=0$, the
green line to $\omega_{re}=\frac{2}{3}$ and the blue line to
$\omega_{re}=1$. The yellow shaded region indicate the Planck 2018
constraint.}}
\end{figure}

\begin{table}[ht]
\caption{The ranges of e-fold number and temperature during
reheating epoch for T-model potential which are consistent with the Planck 2018.}\vspace{0.5cm} % title of Table
\centering % used for centering table
\begin{tabular}{c c c c c} % centered columns (5 columns)
\hline\hline %inserts double horizontal lines
$n=1,\alpha=1$ & $\omega_{eff}=-\frac{1}{3}$ & $\omega_{eff}=0$ & $\omega_{eff}=\frac{2}{3}$ & $\omega_{eff}=1$ \\ [0.5ex] % inserts table
%heading
\hline % inserts single horizontal line
$N_{re}$ & $<8$ & $<20$ & $<52$ & $<27$ \\ % inserting body of the table
$\log(T_{re})$ & $>20$ & $>16$ & $>7.5$ & $>12$ \\ [1ex] % [1ex] adds vertical space
\hline %inserts single line
\end{tabular}
\label{table:nonlin} % is used to refer this table in the text
\end{table}

\subsection{Post-Inflationary era}

The reheating evolution of the Horndeski models is much different
than the usual minimal coupling theory. The scalar field (inflaton)
oscillates very fast and coherently around the minimum of its
potential. The averaged expansion of the universe is given by (for
more details refer to Refs. [34,43,44])

\begin{equation}
\bar{H}(t,\omega_{re})=\frac{1}{\frac{3}{2}(1+\bar{\omega}_{re})(t-t_{end})+H^{-1}_{end}},
\end{equation}

for $t_{end}<t<t_{re}$. The averaged equation of state
($\bar{\omega}_{re}$) after inflationary era is determined by
oscillating behavior of the scalar field about the minimum. The
above expression for $\bar{H}$ imply a different relation for
$\bar{\omega}_{re}$. As a result of Ref. [34], for the non-minimal
derivative models, one can obtain different negative values for the
equation of state. In Fig. (4) we plot time evolution of the Hubble
parameter for the possible values of the equation of state.

On the other hand, during the slow-roll phase we have a dual
description for the tensor-to-scalar ratio in the non-minimal
derivative and usual minimal coupling scenario. However, this
degeneracy breaks by different value of the reheating temperature
and equation of state parameter when the reheating phase is taken
into account.

\begin{figure}
\begin{center}\includegraphics{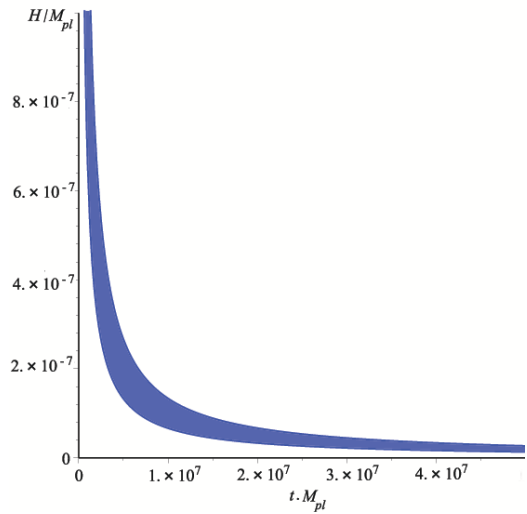} \vspace{6cm}
\end{center}
\caption{\small {Time evolution of the Hubble parameter for the
possible values of the equation of state.}}
\end{figure}

\section{Summary}

In this letter, we have considered a class of non-minimal derivative
inflationary model. We focused on both the E-model and T-model as
generalized models of $\alpha$-attractors. We have obtained the
scaler spectral index and tensor to scaler ratio in the models. As
shown above, in the small $\alpha$ and large $N$ limit, the model
with E-model potential predicts $n_{s}\approx1-\frac{2}{N}$ and
$r\approx\frac{12\alpha}{N^{2}}$ where there is an attractor
behavior. However, the value of the $n_{s}$ and $r$ predicted by
T-model, has some deviation from the corresponding terms in E-model.
We analyzed the parameters numerically for $n=1$ and compared our
results with observations. We have found that the models for large N
are consistent with the Planck data.

Moreover, we have studied the reheating phase in the NMD model. We
obtained e-fold number and temperature during reheating in terms of
scalar spectral index for both E-model and T-model. We also analyzed
the parameters numerically for $n=1$ with $\alpha=1$ and compared
the results with observational data. As shown in Table (1,2), we
have obtained some constraints on the model's parameter space by
adopting our results with Planck 2018.

\end{document}